\documentclass[twocolumn,amssymb,prl,epsfig,superscriptaddress]{revtex4}
\usepackage[utf8]{inputenc}
\usepackage{color}
\usepackage[normalem]{ulem}
 
\usepackage{amssymb}

\usepackage{graphicx}
\usepackage{epsfig}
\usepackage{amssymb}
\usepackage[pdftex, bookmarks,colorlinks=true,urlcolor=blue]{hyperref}

\begin{document}

\title{Effect of  dimensionality on sliding charge density waves. The case  of the quasi-two dimensional TbTe$_3$ system  probed by coherent x-ray diffraction.}
\author{D. Le Bolloc'h}
\affiliation{Laboratoire de Physique des Solides, CNRS, Univ. Paris-Sud, Université Paris-Saclay, 91405 Orsay, France}

\author{A.A. Sinchenko}
\affiliation{Kotel'nikov Institute of Radioengineering and Electronics of RAS, Mokhovaya 11-7, 125009 Moscow, Russia}

\author{ V.L.R. Jacques}
\affiliation{Laboratoire de Physique des Solides, Université Paris-Sud, CNRS, UMR 8502, F-91405 Orsay, France}

\author{L. Ortega}
\affiliation{Laboratoire de Physique des Solides, Université Paris-Sud, CNRS, UMR 8502, F-91405 Orsay, France}

\author{J.E. Lorenzo}
\affiliation{Institute Néel CNRS and Université Grenoble-Alpes, BP166, 38042 Grenoble, France}

\author{G. Chahine}
\affiliation{ESRF, 71 avenue des Martyrs, 38000 Grenoble, France}

\author{P. Lejay}
\affiliation{Institute Néel CNRS and Université Grenoble-Alpes, BP166, 38042 Grenoble, France}

\author{P. Monceau}
\affiliation{Institute Néel CNRS and Université Grenoble-Alpes, BP166, 38042 Grenoble, France}
\begin{abstract}

We report on sliding Charge Density Wave (CDW) in the quasi two-dimensional TbTe$_3$ system probed by coherent x-ray diffraction combined with {\it in-situ} transport measurements. We show that the non-Ohmic conductivity in TbTe$_3$ is made possible thanks to a strong distortion of the CDW. Our diffraction experiment versus current shows first that the CDW remains undeformed below the threshold current I$_S$ and then  suddenly rotates and reorders by motion above threshold. Contrary to quasi-one dimensional systems, the CDW  in  TbTe$_3$  does not display any  phase shifts below  I$_S$  and tolerates only slow spatial variations of the phase above. This is a first observation of  CDW behavior in the bulk in a quasi-two dimensional system allowing collective transport of charges at room temperature.

\end{abstract}

\maketitle

Interaction between pairs of quasiparticles often leads to broken-symmetry ground states in solids. Typical examples are the formation of Cooper pairs in superconductors, charge-density waves (CDWs) and spin-density waves  driven by electron-phonon or electron-electron interactions\cite{gruner_density_2000}. 
The CDW ground state is characterized by a  spatial modulation $\eta \cos(2k_Fx+ \phi)$ of the electron density and a concomitant periodic lattice distortion with the same 2k$_F$ wave vector leading to a gap opening in the electron spectrum. 

The first CDW systems were discovered in the beginning of the 70’s in two-dimensional transition metal dichalcogenides MX$_2$\cite{wilson_charge-density_1975}. CDW state was then discovered in quasi-one dimensional systems like NbSe$_3$, TaS$_3$, the blue bronze K$_{0.3}$MoO$_3$ and in organic compounds like TTF-TNCQ. However, the most remarkable property of a CDW has been discovered a few years later in quasi one-dimensional systems: a CDW may {\it slide} carrying correlated charges\cite{monceau_electronic_2012}.
The sliding mode is achieved when an electric field applied to the sample is larger than a threshold value, manifesting then collective Fröhlich-type transport. This sliding phenomenon is clearly observed by transport measurements. The differential resistance remains constant up to a threshold current and then decreases for larger currents in addition to the generation of an ac voltage, the frequency of which increases with the applied current\cite{monceau_electronic_2012}.

  In spite of numerous studies, the physical mechanism leading to the sliding phenomenon is still far to be fully understood.
One of the difficulties comes from the fact that the sliding mode displays two different aspects. On the one hand, the CDW  is a classical state, similar to an elastic object in presence of disorder\cite{feinberg_elastic_1988}, displaying creep, memory effects and hysteresis\cite{brazovskii_pinning_2004,giamarchi_moving_1996}. On the other hand, a CDW is a macroscopic quantum state\cite{bardeen_theory_1979}, carrying charges by tunneling through disorder\cite{adelman_field-effect_1995} and displaying Aharonov-Bohm effects\cite{latyshev_aharonov-bohm_1997} over microscopic distances\cite{tsubota_aharonov-bohm_2012}.

Recently a new class of quasi-two dimensional CDW compounds, rare-earth tritellurides RTe$_3$, have raised an intense research activity thanks to their peculiar properties\cite{dimasi_chemical_1995,ru_magnetic_2008,gweon_direct_1998}. RTe$_3$ structures  are  orthorhombic ({\it Cmcm}) but the $a$ and $c$ lattice parameters lying in the Te planes  are almost equal (c-a=0.002 \AA\  with $a$=4.307\AA\ for TbTe$_3$ at T=300K) and the double Te-layers  are linked together by a $c$-glide plane. The almost square Te sheets lead to  nearly isotropic properties in the (a,c) plane. The resistance measured along $a$ and $c$ differs by only 10$\%$ at 300K in TbTe$_3$\cite{sinchenko_spontaneous_2014} and the Fermi surface displays an almost square-closed shape in the (a*,c*) plane\cite{brouet_angle-resolved_2008}. 
These quasi-two dimensional systems exhibit a unidirectional CDW wave vector along c* (2k$_F\sim2/7$ c* in TbTe$_3$)  and a surprisingly large Peierls transition temperature, around 300 K, through the whole R-series and above for lighter rare-earth elements. The stabilization of the CDW in TbTe$_3$ over the almost square underlying atomic lattice is reminiscent of copper-oxide planes in high temperature superconductors in which a CDW state was also recently observed\cite{ghiringhelli_long-range_2012}.

However, the most surprising property of TbTe$_3$ is its ability to displays non-linear transport\cite{sinchenko_unidirectional_2014} despite the two-dimensional character of the atomic structure. The aim of the present work is to show that, despite similar resistivity curves, the depining process in quasi one and  two-dimensional systems are quite different. For that purpose, coherent x-ray diffraction has been used  to study the behavior of the 2k$_{F}$  satellite reflection upon application of an external current.

\begin{figure}[!ht]
\includegraphics[width=0.9\columnwidth]{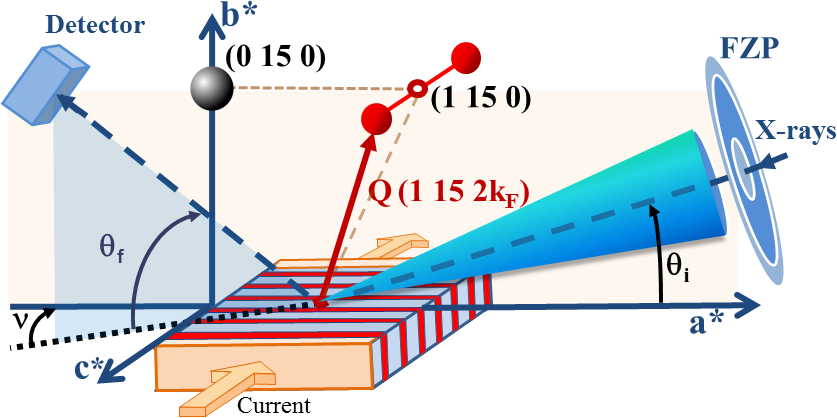}
\caption{a) Experimental diffraction setup (not to scale) with a  coherent $0.5\times 0.5\mu m$ focused x-ray spot. The (1 15 2k$_F$) CDW satellite reflection, associated to the (1 15 0) Bragg (forbidden) Bragg reflection, has been probed with a 2D detector mounted on a lifting detector arm.
}
\end{figure}

As the sliding state of a CDW mainly involves fluctuations of the CDW phase to overcome pinning centers, coherent x-ray diffraction is a suitable technique thanks to its high sensitivity to the phase of any modulation. The extreme case of a single phase shift, such as a dislocation, can locally induce the disappearance of Bragg peaks while a single topological defect is very difficult or impossible to detect with conventional x-ray beams\cite{jacques_bulk_2011}. This approach enabled us  to highlight a dislocation of the CDW\cite{le_bolloch_charge_2005} and an regular array of CDW dislocations\cite{le_bolloch_observation_2008}. 
Thanks to high brilliance of synchrotron sources and improved optics, coherent x-ray beams, tens of micrometers in size, can be obtained with very similar coherence properties to lasers\cite{jacques_estimation_2012}. The experiment described herein was performed at the ID01 beam line of the ESRF synchrotron. A channel-cut Si(111) monochromator has been used which a longitudinal coherence length $\xi_L={\lambda^2}/{2\Delta{\lambda}}=0.6 \mu m$ at E=7.4 keV ($\lambda$=1.675\AA\ ). At this energy, the penetration length of the x-ray beam is $\mu^{-1}\approx 4\mu m$  allowing us to probe the sample volume.
The optical path was defined by a slit opened at $S_0=20(H)\times 60(V)$ $\mu m^2$  at $40cm$ from the sample, followed by a Fresnel zone plate  that focused the coherent x-ray beam down to $0.5 \mu m\times 0.5\mu m$, $23cm$ further on the sample.  The 2D diffraction patterns have been recorded with a pixel detector ($55\mu m\times  55\mu m$  pixels size) and  located at $1.2 m$ from the sample.

\begin{figure}[!ht]
\includegraphics[width=0.9\columnwidth]{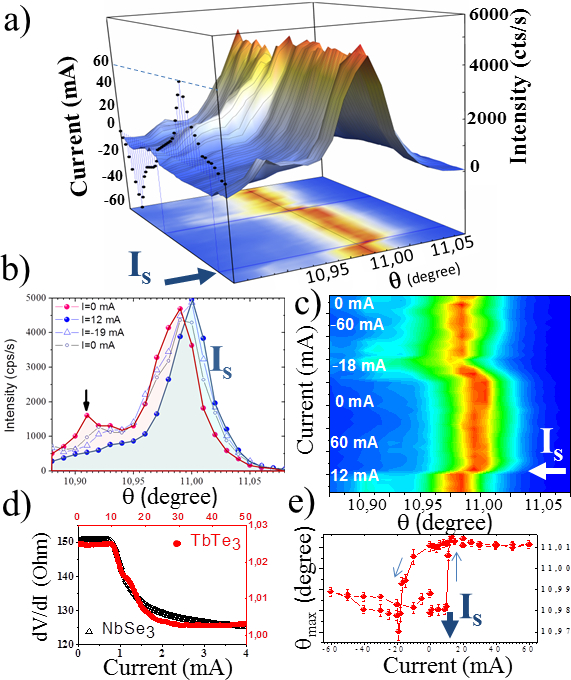}
\caption{a) Rocking curve of the  $Q=(1\ 15\ 2k_F)$ satellite reflection associated to the CDW in $TbTe_3$ versus external current at T= 300K. The intensity corresponds to the sum over the whole pixel camera. The left panel corresponds to the corresponding  currents applied to the sample. b) Intensity profile of the $Q=(1\ 15\ 2k_F)$ satellite reflection for several currents showing the shifted profile at I$_s$ in blue and c) the corresponding 2k$_F$ profile versus current. d) Differential resistance  measured {\it in situ} during the x-ray experiment   in TbTe$_3$ (red dots) and in  $NbSe_3$ (black triangles from \cite{pinsolle_creep_2012}) with $I_S=11 mA$ for TbTe$_3$. e) Main satellite position versus current showing a large hysteresis (a Lorentzian profile has been used to fit the main peak).  }
\end{figure}

A slightly modified  method like the one described in \cite{fischer} was used to the grow high-quality  TbTe$_3$ crystals. 
A $1 mm^2$ square sample, $1.6 \mu m$ thick, was then selected and  cleaved to obtain an elongated shape (1mm long and 120$\mu m$ wide). The c* orientation has been checked by diffraction thanks to the reflection conditions of the {\it Cmcm} space group: (0, k l), k=2n and (h 0 l), h, l=2n. A four-contact method was used for transport measurements with 0.6mm between the two inner contacts. The resistance ratio between  room and  helium temperature  was typically larger than 100, similarly to previous reports\cite{ru_magnetic_2008}. The current-voltage curve has been regularly measured during the experiment, showing very stable threshold current $I_S=11 mA$ (Fig.2d). The  threshold remained stable during several days of experiment suggesting no radiation damage by x-rays. This conclusion is reinforced by the remarkable stability  of the  2k$_{F}$ satellite intensity and of the resistance during acquisitions. This is an important difference with  quasi one-dimensional systems  such as NbSe$_3$ and K$_{0.3}$MoO$_3$ in which the CDW is irreversibly damaged by too intense x-ray beams.

\begin{figure}[!ht]
\includegraphics[width=0.9\columnwidth]{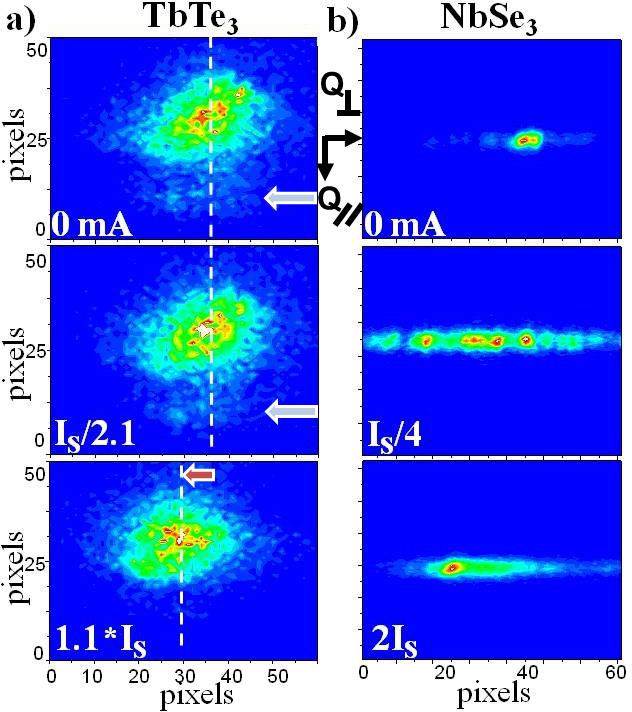}
\caption{Coherent diffraction patterns of the $2k_F$ satellite reflection associated to the CDW versus external current, below and above the threshold current I$_S$, in a) the quasi-two dimensional TbTe$_3$ system (for I=0mA, I=I$_S$/2.1=5mA and I=1.1$\times$ I$_S$=12mA) and  b) in the quasi-one dimensional NbSe$_3$ system (from \cite{pinsolle_creep_2012}). The 2D images are a sum over several $\theta$ angles through the maximum of intensity.  For TbTe$_3$, the blue arrow indicates the contribution of the disoriented domain contributing in the profile in Fig.2b and the red one indicates the shift of the 2k$_F$ reflection at I$_S$. Although the cutting plane is different in the two cases, the vertical  direction of the camera  is close to the 2k$_F$ wave vector  (Q$_\parallel$) and the horizontal one is transverse to  2k$_F$ (Q$_\perp$) in both cases.}
\end{figure}

The  Q(1 15 2k$_F$) satellite reflection associated to the CDW has been measured in reflection geometry at room temperature (see the setup in Fig.1 with  $\theta_i\approx11^o$, $\theta_f=63.8^o$ and $\nu=12.9^o$).
 A micrometer  $0.5 \mu m\times 0.5\mu m$ coherent beam has been used, focused with Fresnel zone plates (FZP).
In the pristine state, without external current, the sample displays a transverse CDW correlation lengths of $\xi_T\approx 40 nm$ (obtained from the main peak of the rocking curve in Fig.2b). The rocking curve of the 2k$_{F}$ satellite displays  also a second contribution from a disoriented domain, three times weaker in intensity (see the arrow in Fig. 2b and Fig. 3a). The coherent diffraction pattern displays many speckles distributed almost isotropically and coming from CDW phase shifts in the ({\bf a}*,{\bf c}*) plane and between sheets along {\bf b}*-direction (see Fig. 3a). 

The scattering features of the Q(1 15 2k$_F$) satellite reflection were studied with respect to  applied currents, both below and above the threshold current  (I$_S$=11mA). Typical excursion ranged from I=0mA up to 60mA and then back to I=0mA (see Fig.2a and 2c).
The increase of current from I= 0mA to below the threshold current hardly changes the diffraction pattern. The distribution of speckles does not change although  slight variations in intensity are observed  (see Fig. 3a).

At I$_S$,  a drop of 2\% of the differential resistance is measured and the satellite reflection changes. The satellite's position  increases by 0.013$^o$ in $\theta$  (see the blue curve in Fig.2b), the width of the main peak decreases, its intensity increases slightly and the small contribution coming from the disoriented domain  disappears.  Just above the threshold current, an increase of  14\% of the  transverse CDW correlation length $\xi_T$ is observed. The isotropic diffraction pattern remains but the distribution of speckles is totally different.
The CDW reorders above I$_S$ but also rotates by an angle $\beta$=0.02$^o$  (see the red arrow in Fig.3a) with a complete reorganization of domains.

The rotation of the wave vector is clearly observed when using a very small $0.5\mu m$ beam. When  the measurement is performed with a  100 times larger beam (without focusing with the FZP), averaging over a 100$^2$ times larger volume,  the 2k$_F$ rotation  is still observed but significantly less pronounced. Increasing the spatial average blurs the signature by including contributions of smaller domains.

The CDW behavior versus current is almost reversible but with a strong hysteresis (see Fig. 2e).
The gradual increase in current from I$>$I$_S$ to I=5.45$\times$I$_S$=60mA induces almost no change compared to the state just above the threshold current. The satellite remains also unchanged when decreasing current down to I=0mA and even further down to negative currents. Only at I =-19mA does the satellite resumes its original state. Back to I = 0mA finally, the 2k$_F$ profile is close to the initial state, but not identical however, with a larger width and smaller intensity. The  contribution from the small disoriented domain also reappears (see Fig 2b). All these observations show that the distorted CDW state  is a frozen metastable state which can be released by reversing the applied current.

The experiment was repeated several times. When changing the beam position on the sample, the CDW displays  similar features, including  rotation and  narrowing. However the rotation direction and its amplitude may change.
The fundamental Bragg peaks $q_0$ does not depend on current (within our resolution) excluding any thermal effect induced by  external currents. 

This measurement first demonstrates that the CDW deformation is directly involved in the non-linear transport observed in TbTe$_3$. The sliding state  is reached thanks to a  rotation of the 2k$_F$ wave vector to overcome pinning centers and a reordering of the CDW  by motion. This rotation versus current is almost reversible and strongly hysteretic. Motional narrowing  and hysteresis are common features with quasi-one dimensional systems like NbSe$_3$\cite{pinsolle_creep_2012} and K$_{0.3}$MoO$_3$\cite{le_bolloch_observation_2008,jacques_evolution_2012}.

However, the two types of systems widely differ from each other by the type of CDW distortions. Contrary to TbTe$_3$,  NbSe$_3$ displays creep for  currents well below the threshold. This  corresponds to the presence of speckles along a line in the reciprocal lattice (see Fig. 3b) due to abrupt CDW phase shifts parallel to the chain axis in real space (see Fig. 4d)\cite{pinsolle_creep_2012}. 
Rotation of the 2k$_F$ wave vector  is not observed in NbSe$_3$ while creep is not observed in TbTe$_3$. 

\begin{figure}[!ht]
\includegraphics[width=0.9\columnwidth]{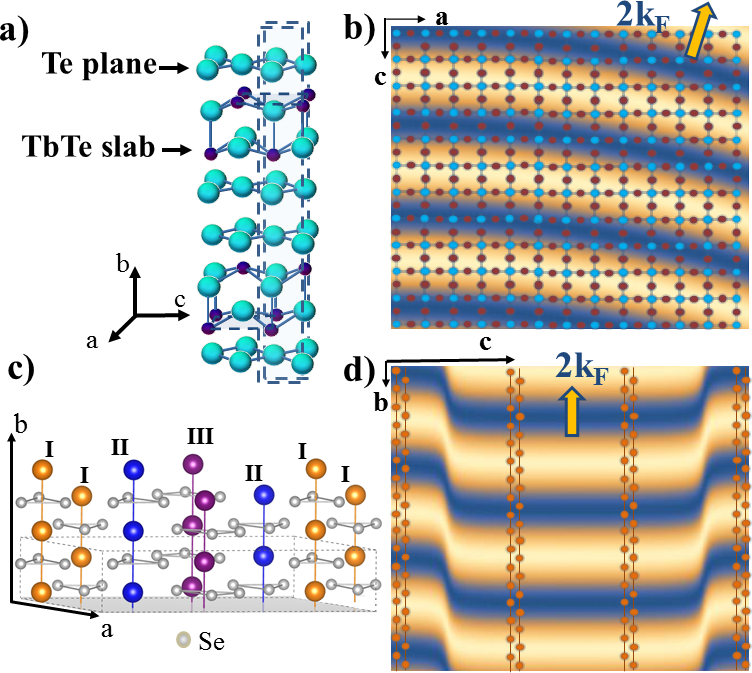}
\caption{a) Sketch of the TbTe$_3$ structure and b) its projection along $b$ in the ($a,c$) plane. c)  NbSe$_3$ structure with the three types of chains (only the first chain, the orange one, participates to the $q_1$ CDW) and d)  projection  of  chains I  along $a$ in the ($b,c$) plane. In  both systems, the CDW is represented by blue wave fronts in b) and d).}
\end{figure}

This difference in behavior is mainly explained by the difference of dimensionality of the two systems.
Contrary to the quasi two-dimensional TbTe$_3$ system,  NbSe$_3$  is quasi one-dimensional, made of parallel chains of atoms along which the resistance and the elastic constants are at least 10 times smaller than in the two  other perpendicular directions.   TbTe$_3$ displays an almost square Fermi surface, while NbSe$_3$ displays an open Fermi surface, consisting of two nearly parallel planar surfaces perpendicular to the atomic chains\cite{schafer_unusual_2003}.

The TbTe$_3$ behavior above I$_S$ can be explained by elastic theory, considering the CDW as an incompressible lamellar phase in the presence of disorder. We take into account the sliding state  and consider only spatial  fluctuations of the phase $\Phi(x,y)$\cite{feinberg_elastic_1988} where $x$ is the direction parallel to 2k$_F$.  Transverse deformations which do not compress or dilate the CDW modulation are neutral and will be favored, thus: $k_a\frac{\partial^2\Phi}{\partial^2 y}=qE$ where $k_a$ is the elastic constant  along the transverse $a$-axis and $E$ is the applied field. 
In  TbTe$_3$, the underlying lattice seen by the CDW is almost homogeneous and k$_a$ can be considered constant in space. The solution is thus parabolic: $\Phi(y)\propto y(l-y)$. If $l$ is a macroscopic distance corresponding to distances between strong pinning centers,  side edges or  surface steps, $\Phi(y)$ slowly varies with space, inducing a rotation of the 2k$_F$ wave vector as shown in Fig.4b, in agreement with our measurement.

In NbSe$_3$, only one type of Nb chain out of three is involved in the q$_1$ CDW (called chains I in orange  in the structure displayed in Fig.4c). Since more than 15.6\AA\  separates two groups of chains I, the effective elastic constant strongly varies along the transverse direction and the  continuous elastic model used for TbTe$_3$ is not appropriate anymore. This strong structural anisotropy leads to abrupt CDW phase shifts  as illustrated in Fig.4d (without compressing or expanding CDW wave fronts).  In that case, speckles appear along a line, without tilt, in agreement with our measurement.

Below threshold, the two systems also display a very different behavior. At low currents, the strong deformation of the CDW  in NbSe$_3$ is not observed  in TbTe$_3$ in which  the diffraction pattern remains nearly identical with the same distribution of  speckles up to I$_S$. Contrary to NbSe$_3$ and K$_{0.3}$MoO$_3$, the homogeneity of the TbTe$_3$ atomic lattice seems to prohibit all types of  CDW wave fronts deformation, suggesting a strong pinning scenario\cite{fukuyama_dynamics_1978} due to dimensionality. 

At I$_S$, the CDW depins abruptly and hysteretically as expected from strong pinning theory\cite{tucker} where the phase adjusts locally around each pinning centers for I$<$I$_S$.
Note that the sudden  collective depinning  at I$_s$, measured by diffraction which probes sub-micrometer domains, is totally correlated with transport measurements averaging over the 0.6mm gap between the two electrodes. In the sliding state however, only slow space variations of the phase are observed, suggesting a collective deformation of the phase over large distances. This sudden CDW deformation at the threshold current, without creep below threshold, may be related to the strong electron-phonon coupling  invoked to explain the soft mode behavior  in TbTe$_3$\cite{maschek}.

\end{document}